\documentclass[manuscript]{acmart}

\usepackage{url}
\AtBeginDocument{%
  }

\setcopyright{acmlicensed}
\copyrightyear{2018}
\acmYear{2018}
\acmDOI{XXXXXXX.XXXXXXX}

\acmJournal{JACM}
\acmVolume{37}
\acmNumber{4}
\acmArticle{111}
\acmMonth{8}

\begin{document}

\title{Photographic Conviviality: A Synchronic and Symbiotic Photographic Experience through a Body Paint Workshop}

\author{Chinatsu Ozawa}
\affiliation{%
  \institution{Digital Nature Group, University of Tsukuba}
  \city{Tsukuba}
  \country{Japan}}
\email{toremolo72@digitalnature.slis.tsukuba.ac.jp}

\author{Tatsuya Minagawa}
\affiliation{%
  \institution{\mbox{R\&D Center for Digital Nature}}
  \city{Tsukuba}
  \country{Japan}}
\email{mina.tatsu@digitalnature.slis.tsukuba.ac.jp}

\author{Yoichi Ochiai}
\affiliation{%
  \institution{\mbox{R\&D Center for Digital Nature}}
  \city{Tsukuba}
  \country{Japan}}
\email{wizard@slis.tsukuba.ac.jp}

\renewcommand{\shortauthors}{Ozawa et al.}

\begin{abstract}
This study explores "Photo Tattooing," merging photography and body ornamentation, and introduces the concept of "Photographic Conviviality." Using our instant camera that prints images onto mesh screens for immediate body art, we examine how this integration affects personal expression and challenges traditional photography. Workshops revealed that this fusion redefines photography's role, fostering intimacy and shared experiences, and opens new avenues for self-expression by transforming static images into dynamic, corporeal experiences.
\end{abstract}




\maketitle

\begin{figure*}[ht]
    \centering
    \includegraphics[width=1.0\linewidth]{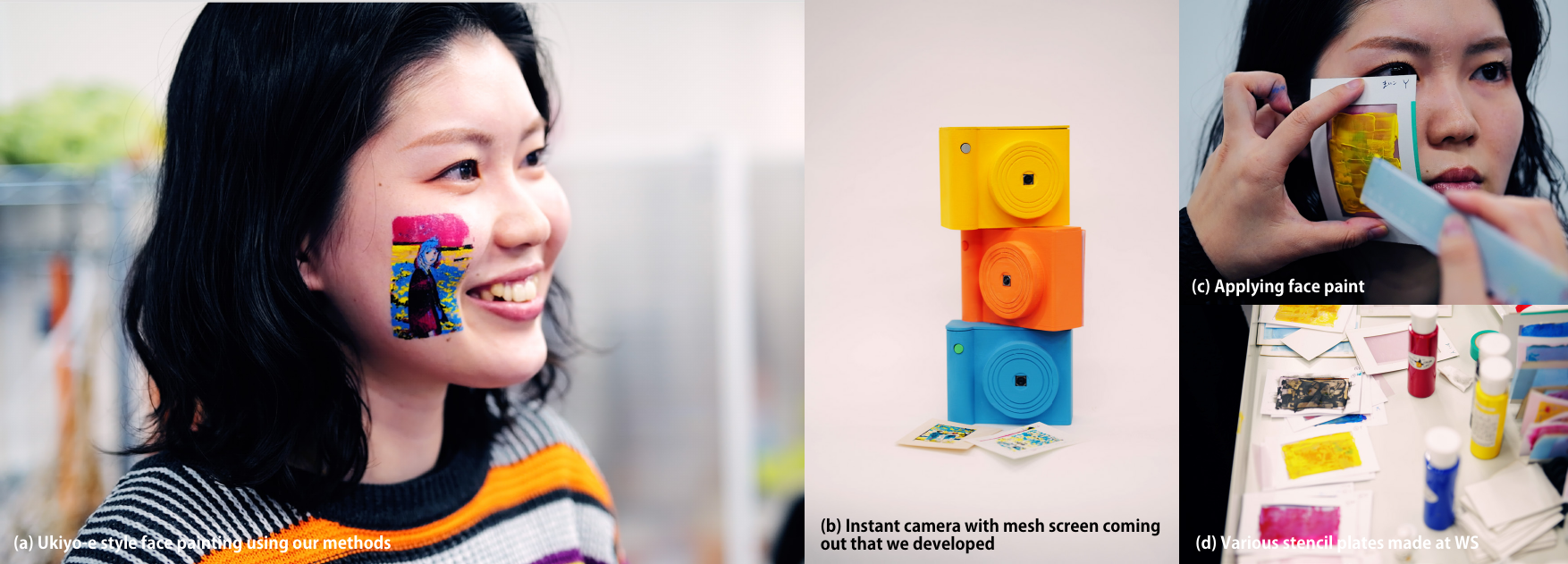}
    \caption{(a) A woman applying her portrait to her cheek using our method. (b) The camera we developed, in three colors matching the output images. (c) The body painting process. (d) A workbench with the produced meshes.}
    \label{fig:teaser}
\end{figure*}

\section{Introduction}
Since photography's beginning, artists have reimagined its relationship with art, seeking new expressions. For instance, Gerhard Richter blurred the boundaries by imitating a photograph with a painting. “Photo painting", known as Richter’s way, fuses photographic objectivity with painting subjectivity and creates a new trend.
Similarly, body ornamentation has been influenced by photography. As painting expanded through photography, body ornamentation embraced new possibilities via photographic expressions. Tattoo artist Scott Campbell uses precise lasers to etch realistic designs onto skin, mirroring photos. He also creates three-dimensional works that bring tattoo culture into contemporary art, using tattoos as motifs. This approach, denominating “photo tattooing” as inspired by photo painting, is a theme not yet fully explored in media art. This paper examines how photographic technology and culture impact body ornamentation.
Considering body ornamentation's influence on photography, if snapshots could instantly decorate the body like tattoos, how might this change relationships with photography and self? Traditional tattoos involved artisans inscribing images onto skin using machines. Now, instant cameras with silk-screen printing can immediately embellish the body with photos. What thoughts and feelings come? This exploration seeks new potentials of photography and introduces new cultural critiques.
This research reevaluates the border between photography and body ornamentation, exploring new expressions and communication from their interplay. This paper makes the following contributions:

\begin{enumerate}
    \item New lens for art-historical interpretation by surveying digital art in photography and tattooing history.
    \item Case studies and discussions of a workshop with instant cameras we proposed in body painting and photo tattooing.
\end{enumerate}

Through these endeavors, the aim is to illuminate how the evolving relationship between photography and humanity expands self-expression's horizons.

\section{The Fusion and Evolution of Photographic Technology and Body Ornamentation}

Tattooing is one of the oldest forms of body modification practiced worldwide since ancient times. The European Iceman had 61 linear tattoos, and the earliest figurative tattoos have been found on mummies from ancient Egypt\footnote{https://www.nationalgeographic.com/history/article/131016-otzi-ice-man-mummy-five-facts}.
Globally, painting developed around religious symbols, myths, and geometric patterns. In Europe, Christian religious paintings and myths were common subjects. Tattoo motifs were primarily religious symbols and geometric patterns, emphasizing realism and religious significance.
In contrast, during Japan's Edo period, ukiyo-e art featured a variety of subjects such as everyday life, customs, kabuki actors, portraits of beauty, and landscapes. Bold lines, vivid colors, and flat compositions characterized these works. This aesthetic, known as Japonism, influenced Western artists ~\cite{maki2014connections}. The expressions in ukiyo-e also influenced tattoo culture, incorporating warriors, mythical figures, and natural landscapes into tattoo designs. The grand, highly artistic Japanese tattoos that cover the entire body are recognized as a unique body ornamentation culture distinct from Western tattoos.
This chapter provides an overview of digital art in the history of photography and tattooing, offering a new perspective.

\subsection{Japanese Tattoos and Photography}

\begin{figure}[htbp]
    \centering
    \includegraphics[width=1.0\linewidth]{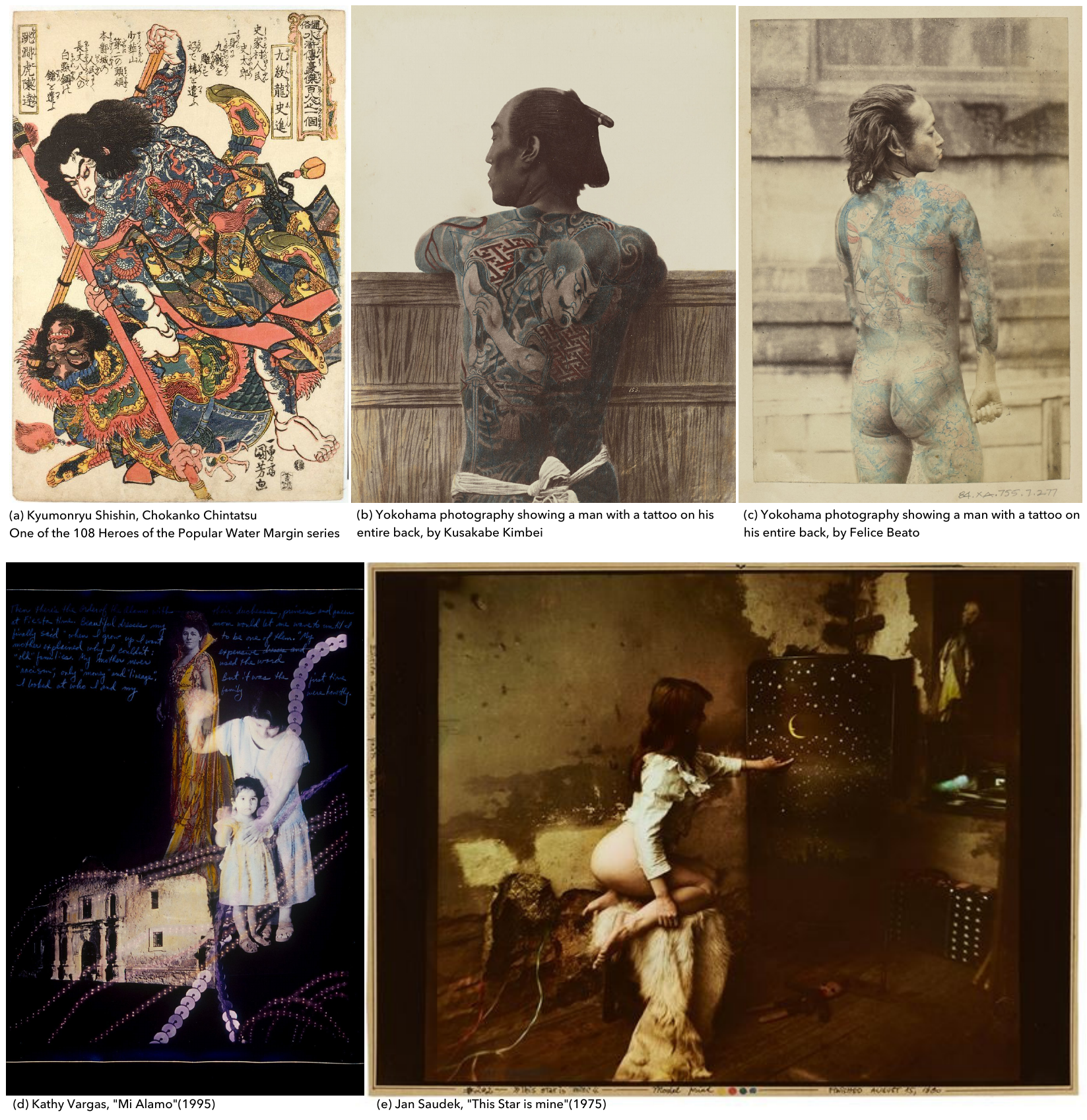}
    \caption{Fusion of photography and coloring: (a) Utagawa Kuniyoshi's "Water Margin" inspired ukiyo-e with heroic figures bearing full-back tattoos, influencing people like (b) firefighters and rickshaw pullers. (c) Beato's "Yokohama Photography" used hand-colored photos with Japanese aesthetics. Contemporary artists like (d) Kathy Vargas ("Mi Alamo") and (e) Jan Saudek ("This Star is Mine") applied hand-coloring techniques, expanding photographic expression in art.}
    \label{fig:ukiyoe}
\end{figure}

Japan's tattoo culture is ancient, with traces found in Jomon period clay figurines and incised pottery and haniwa from the Yayoi and Kofun periods~\cite{maeda_jomon_tattoo}. "Kojiki" and "Nihon Shoki" mention tattoos as customs of peripheral peoples or as punishment~\cite{nihon_shoki}~\cite{kojiki}. Tattoos were used to inscribe patterns on the body as vows to lovers or deities, or as punishment.
Around the Meiwa period of the Edo era (1764–1772), tattooing emerged as a form of pictorial ornamentation, with intricate multicolored designs carved primarily on the back. "Bunshin Hyakushi" praises these tattoos as art to be proud of worldwide~\cite{miyashita_art_history}.
Ukiyo-e artist Utagawa Kuniyoshi produced many works based on "The Water Margin"~\cite{van1982irezumi}(Figure\ref{fig:ukiyoe}(a)). The images of heroes adorned with full-back tattoos fascinated the public. Emulating them, firefighters and rickshaw pullers acquired tattoos (Japanese-style tattoos), and techniques such as "bokashi"(gradation or shading) were developed~\cite{schoneveld2018representation}. With the addition of color, tattoos became established as a sophisticated form of body ornamentation possessing high artistic value.
In the Meiji era, photography was introduced from the West, impacting tattoo culture. Photographer Felice Beato established a studio in Yokohama and photographed Japanese customs and landscapes using hand-colored photographs known as "Yokohama Shashin"~\cite{lacoste2010felice}(Figure\ref{fig:ukiyoe}(b)(c)). His studio included painters and assistants and holds an important place in the history of Japanese photography.
In these hand-colored photographs, tattoos covering the entire backs of Japanese subjects became ideal motifs. By combining realistic photography with color, the detailed designs and colors of the tattoos were faithfully reproduced. Consequently, tattoos were elevated to comprehensive artistic expressions and gained international popularity. These photographs were sold as postcards and albums, becoming important media for introducing Japanese culture to Western audiences~\cite{leyshon_19th_photographs}.

In other countries, hand-colored photography started around 1840, utilizing techniques of coloring photographs with paints(Figure\ref{fig:ukiyoe}(d)(e)). Pioneers such as Antoine Claudet and Johann Baptist Isenring led this movement, and the fusion of photography and coloring spread worldwide~\cite{machado2010politics}. However, hand-colored photographs focusing on tattoos as the main subject were unique to Japan, indicating the high artistic value of Japanese tattoos. In this way, tattoos became connected with photography; their artistry was reevaluated, promoting their evolution and diversification.

\subsection{The Evolution of Digital Data and Body Ornamentation}

\begin{figure}[htbp]
    \centering
    \includegraphics[width=1.0\linewidth]{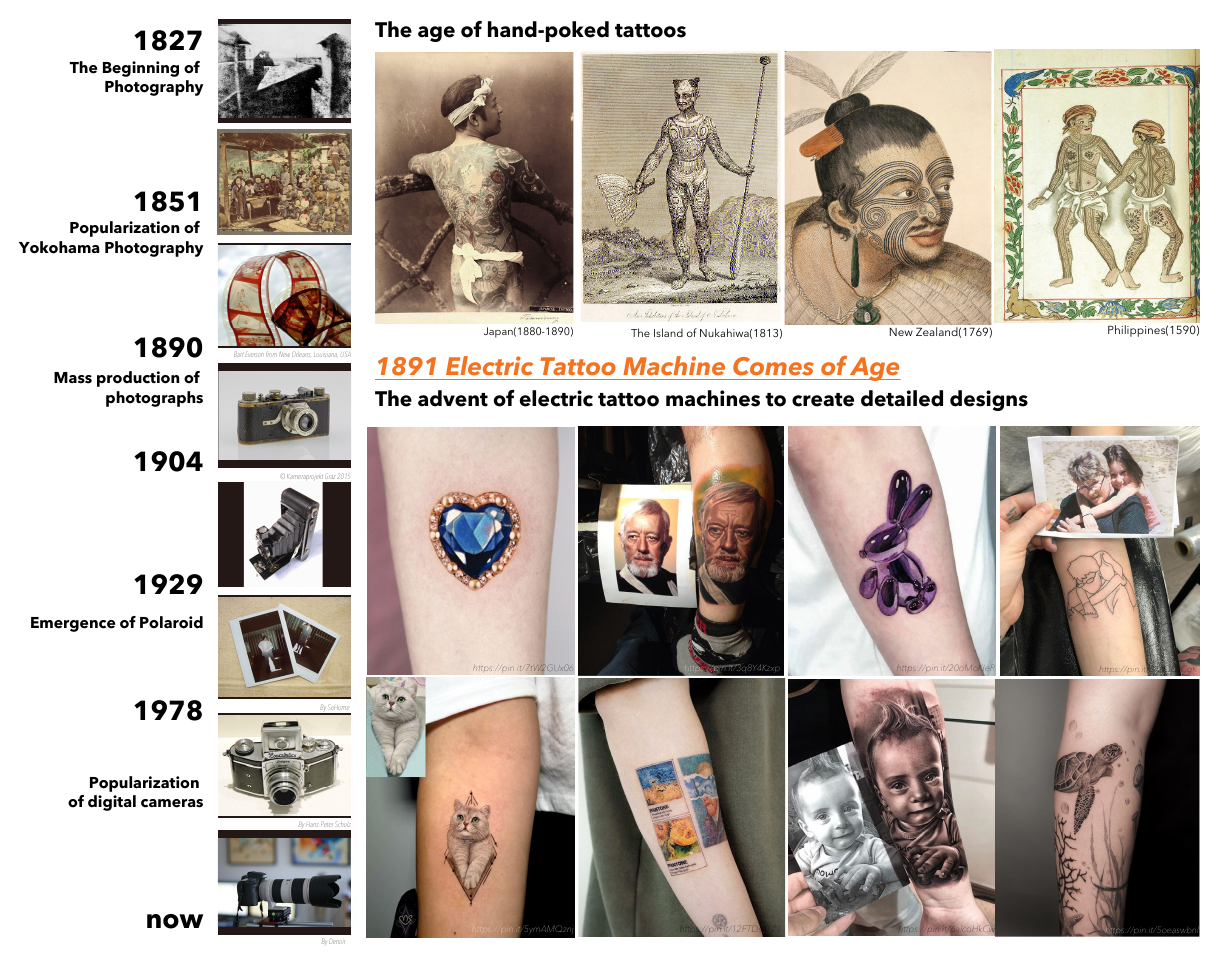}
    \caption{The Development and Transition of Photography and Tattoos}
    \label{fig:tatto-photo}
\end{figure}

In contemporary times, advancements in digital technology have enabled the use of photographic data directly as tattoo designs. The integration of sophisticated tattoo machines with digital technology allows intricate designs to be faithfully reproduced on the skin. As a result, the practice of engraving pixelated data onto the body has become more prevalent.
Contemporary tattoo motifs have diversified. Famous paintings are being tattooed with photographic precision. Designs now include scenes from anime, landscape paintings, and those with three-dimensional visual effects(Figure\ref{fig:tatto-photo}). Achieving these is challenging without using photographs or images. Individuality has also increased. People can bring in photographs to have images of loved ones or pets tattooed, reproducing even fine details such as fur textures and colors.
Thanks to the widespread availability of photography, anyone can easily take photographs and use them as the basis for tattoos. Tattoo artists obtain design inspiration from social media~\cite{ryan2022tattooing}, and it is common for consumers to search for designs and artists online~\cite{silva2022tattoos}. In certain regions, tattoos incorporating family photos are valued as a means to deepen familial bonds~\cite{krtalic2021tattoos}.
Advancements in tattoo machines have made intricate designs possible, and sharing on social media has expanded the range of available designs. Tattoos on areas like the neck and hands have become common, contributing to the spread of a visually oriented-tattoo culture.
In this way, the transition of tattoo motifs is closely linked to the development of photographic technology. Pixel data, which can easily disappear in the digital realm, transforms into permanent information when inscribed on the body. Pixels become special entities symbolizing personal thoughts and emotions. This endows an individual's bodily history with an auratic authority.
As tattoo culture spreads, interest in body painting is also growing. Body painting is convenient, allowing pixels to be immediately applied to the body, and its temporary nature reduces resistance. Many people use photos posted on social media to express themselves, and non-permanent body art serves as a means of self-expression. Especially on Instagram, creativity is emphasized, resulting in many artistic designs.
While tattoos and body painting differ in permanence, incorporating easily lost pixels into the body extends their period of existence. Ephemeral digital data becomes a temporary physical presence through the body, prolonging the time it is observed and encountered by others.

\subsection{Rethinking Photography and Physicality}

\begin{figure}[htbp]
    \centering
    \includegraphics[width=0.9\linewidth]{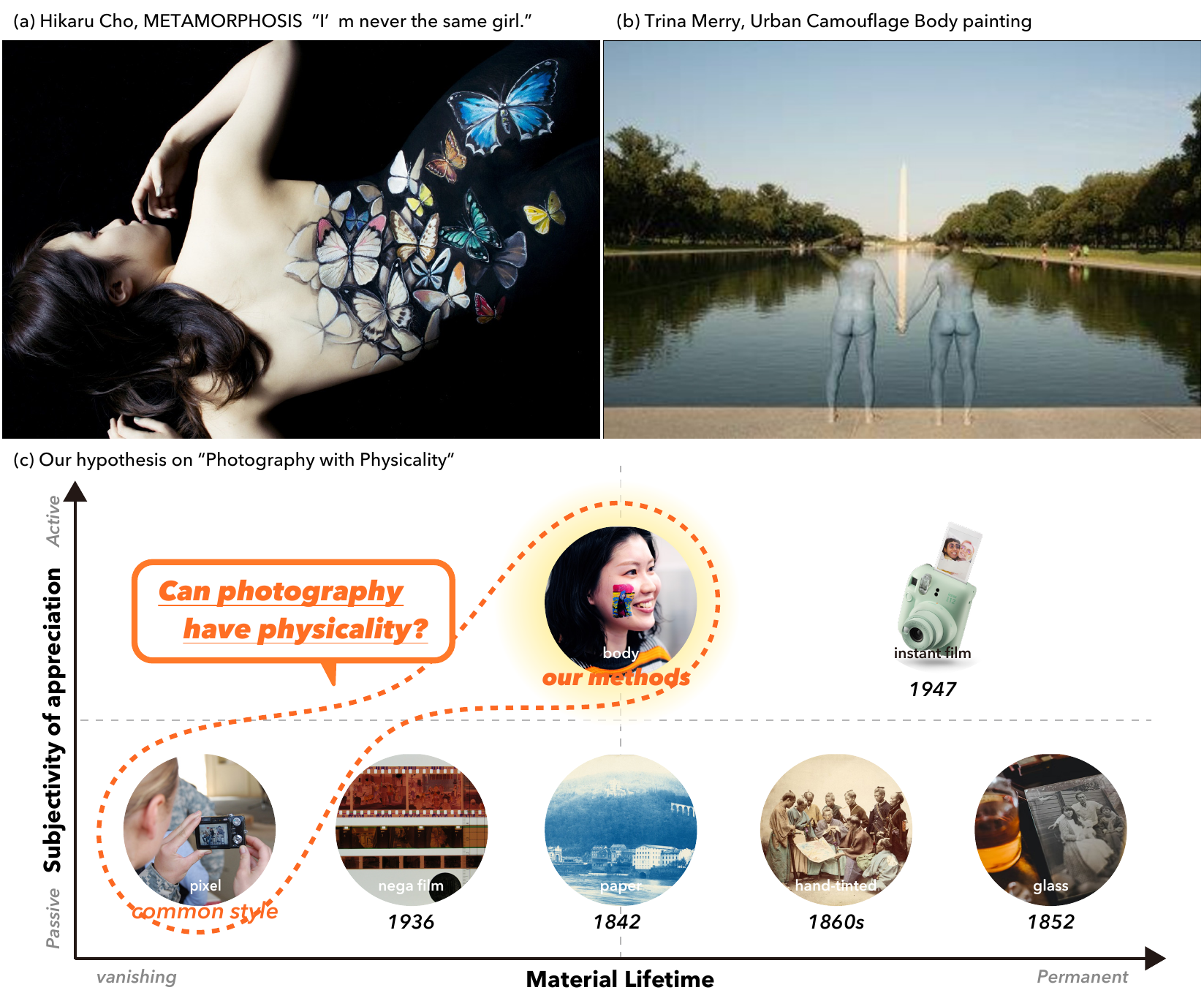}
    \caption{Artists camouflage their bodies within natural landscapes, integrating into photographs. The final artwork is the photograph, and effects vary with background color. Hikaru Cho's works draw motifs on a deep black base; photographing against black creates an illusion where motifs stand out and the body seems translucent.}
    \label{fig:bodyart}
\end{figure}

Photography is the most realistic recording medium and can convey messages and intentions on a single plane. In art, photography has become not only a record of events but also a work of art in itself~\cite{cotton2004photograph}.

While photography was primarily exhibited vertically on walls, Wolfgang Tillmans experimented with discrete arrangements and horizontal displays~\cite{tillmans2006wolfgang}. However, the relationship between the viewer and the photograph remained static and unidirectional.
In contemporary art, installation-like temporary exhibits are becoming more common. Body art has become featured, including Orlan's "La Réincarnation de saint Orlan"~\footnote{\url{https://www.orlan.eu/works/performance-2/}}, Stelarc's "Ear on Arm"~\footnote{\url{http://stelarc.org/_activity-20242.php}}, and Yayoi Kusama's body paintings~\footnote{\url{https://play.qagoma.qld.gov.au/looknowseeforever/essays/performing-the-body/}}.

Photographs are easy to take, artists began recording performances through photographs and using them as artworks. This made viewers indirectly performance witness, and photography played an expansive role in body expression. In recent years, artists introduced in Figure\ref{fig:bodyart}(a)(b) have been exploring new possibilities in body expression. They apply realistic paintings to their bodies and present their works through photography. These are temporary artworks that involve painting on an impermanent medium like the human body, with the paintings eventually fading away. However, through photography, viewers can experience and share their aesthetics. Therefore, body art is an art form that cannot exist without photography.

Thus far, the discussion has focused on how body expression is extended through photography, but it is necessary to reconsider the physicality of photography itself(Figure \ref{fig:bodyart}(c)). Physicality is defined as the phenomena and actions that the body interacts with the external world.

By considering the human body as a dynamic medium with mass, we think about the situation of photography instantly melting into the skin. When photographs are printed on the body, they are transformed into static visual images to physical interactive media. Through the body as the organic interface, photography acquires both input and output functions, and interplay with the external environment.

If the practice of melting digital motifs in the body as tattoos and body paintings becomes widespread, the relationship between photography and humans will deepen further. Ephemeral digital data gain permanence and presence through the body. This leads to redefining the meaning of photography.

Thus, through body ornamentation with instant photography printing, photography attains physicality, and a new relationship between humans and photography is constructed. This evolution, resulting from the fusion of photographic technology and body ornamentation, greatly expands the possibilities of self-expression.

In Chapter 3, the possibility of photography with physicality will be examined through a case study employing a self-made instant camera.

\section{An Instant Camera That Outputs Screen Mesh}

In this study, to demonstrate the concept presented in Section 2.3, we propose an instant camera that outputs screen mesh for silkscreen printing. When a photograph is taken with this instant camera, it immediately processes the image and outputs the mesh screen. By using this mesh as a stencil, multicolor silkscreen printing becomes possible. Detailed explanations of the hardware design in Figure\ref{fig:camera} and image processing methods in Figure\ref{fig:result-variation} are summarized in the Supplemental Document. In this chapter, we introduce the final printing results applied to the skin.

\begin{figure}[htbp]
    \centering
    \includegraphics[width=1.0\linewidth]{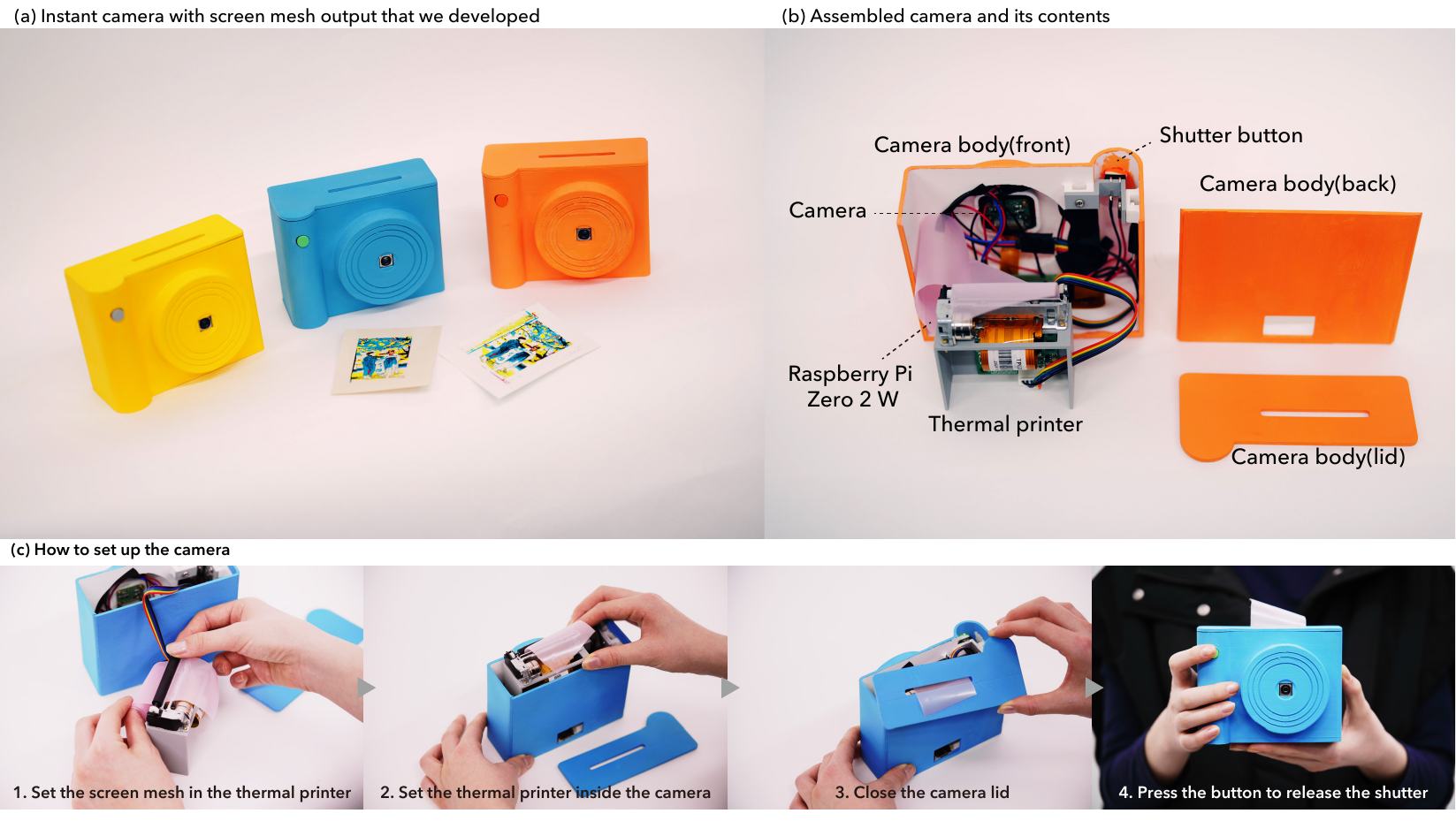}
    \caption{System components: compact thermal printer, Raspberry Pi Zero 2 W, Camera Module v3, and 3D-printed casing. The Raspberry Pi handles image capture, server communication, and printer control. The camera captures images sent to the server. The casing is 3D-printed.}
    \label{fig:camera}
\end{figure}

\begin{figure}[htbp]
    \centering
    \includegraphics[width=1.0\linewidth]{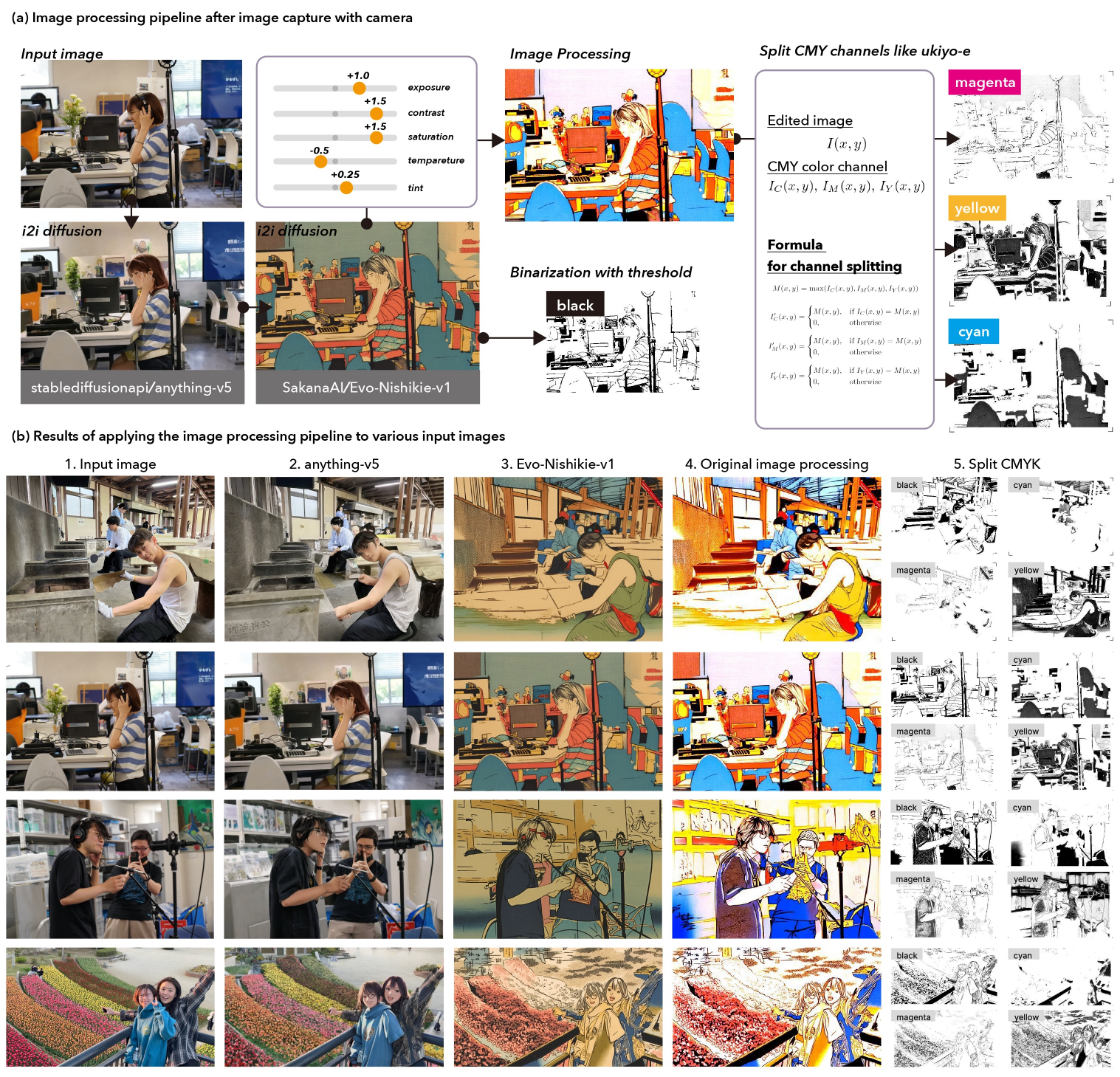}
    \caption{Image processing pipeline converting a full-color photo into ukiyo-e style for silkscreen printing: (1) Acquire user's photo. (2) Convert to illustration with "Anything v5" model. (3) Apply "Evo-Nishikie-v1" model for ukiyo-e style. (4) Perform color correction to remove brown background and enhance colors. (5) Split into CMY color channels, prevent overlap. (6) Extract black layer via binarization for contours and shading. (7) Output four monochrome images for CMYK, used as stencils in silkscreen printing.}
    \label{fig:result-variation}
\end{figure}

\begin{figure}[htbp]
    \centering
    \includegraphics[width=1.0\linewidth]{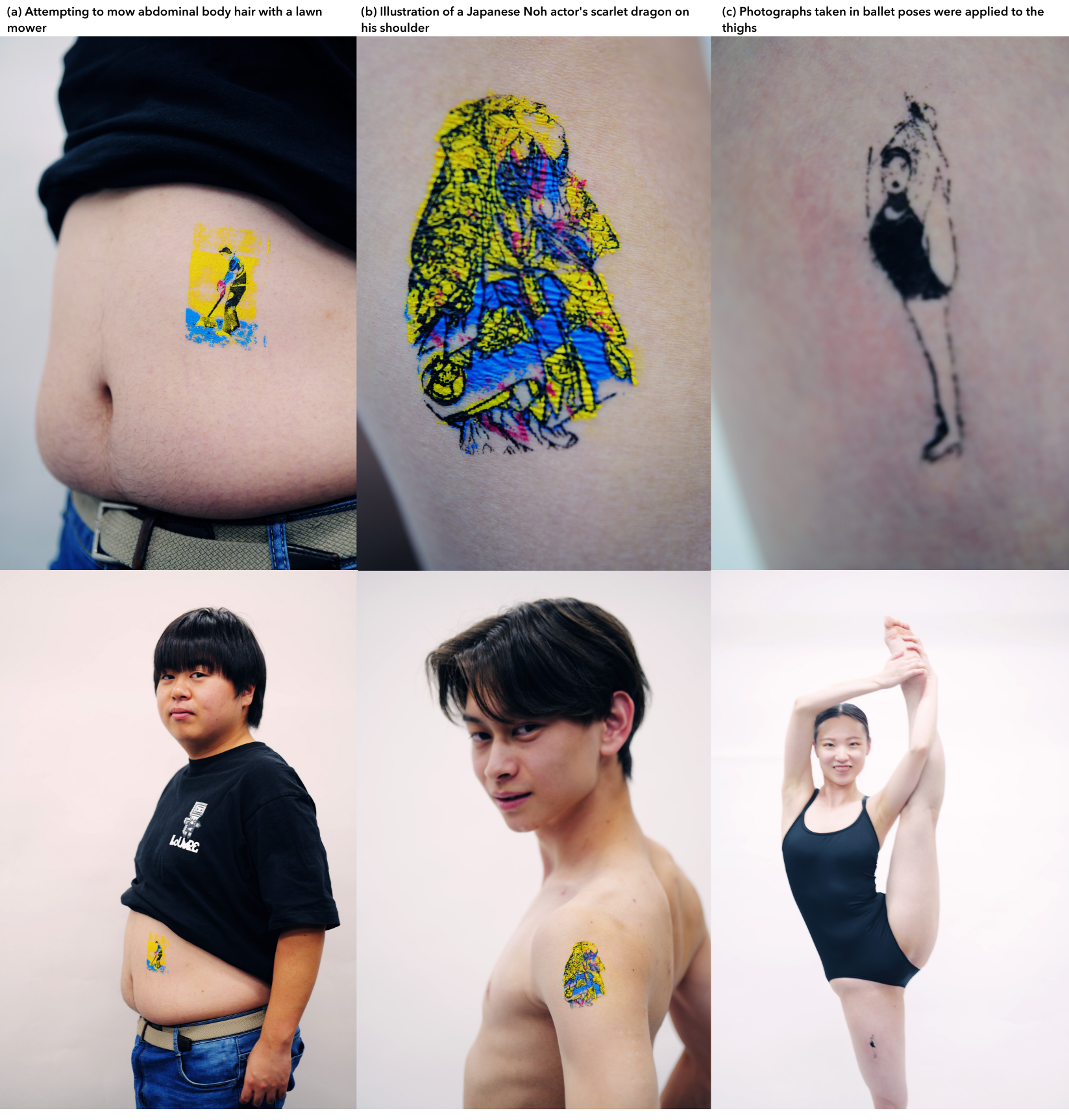}
    \caption{Printing results: Adjusting illustrations to match body pose allows flexible designs. (a) An elderly man with a grass cutter printed to appear as if trimming body hair. (b) A Noh theater dragon printed on the shoulder. (c) A ballet dancer's pose printed on her thigh.}
    \label{fig:results-print}
\end{figure}

Using the four stencils created through our camera, a four-layer silkscreen print is applied to the skin. The outputted mesh screens are affixed to a base. Using these, as shown in Figure\ref{fig:results-print}(a), printing is performed in the order of cyan, magenta, yellow, and black.
Printing is possible even if not following the aforementioned order. For example, printing in the order of the largest print area can prevent misalignment of the stencils. However, since black serves as both contour lines and shadows, printing it last sharpens the image.
Moreover, adding an extra step during printing can further enhance the quality of the image.
For instance, by removing the background colors that extend beyond the black contour lines after printing, leaving only the objects, a more impactful body painting can be achieved. In Figure\ref{fig:results-print}(b), after printing using the basic four-color method, the parts extending beyond the black contours were removed. Depending on the photographed subject, it may also be effective to create a monochrome silhouette. In Figure\ref{fig:results-print}(c), all four layers were printed in black, and the background parts extending beyond the object's contours were removed to create a black-and-white silhouette art.

\section{Workshop}

In this study, a body painting workshop using a self-made instant camera was conducted to explore new expressions arising from the fusion of photographic technology and body ornamentation. Participants were members of the general public recruited in advance. After the workshop, a questionnaire was administered, and semi-structured interviews were conducted with those who wished to participate.

\subsection{Overview}

\begin{figure}[htbp]
    \centering
    \includegraphics[width=0.9\linewidth]{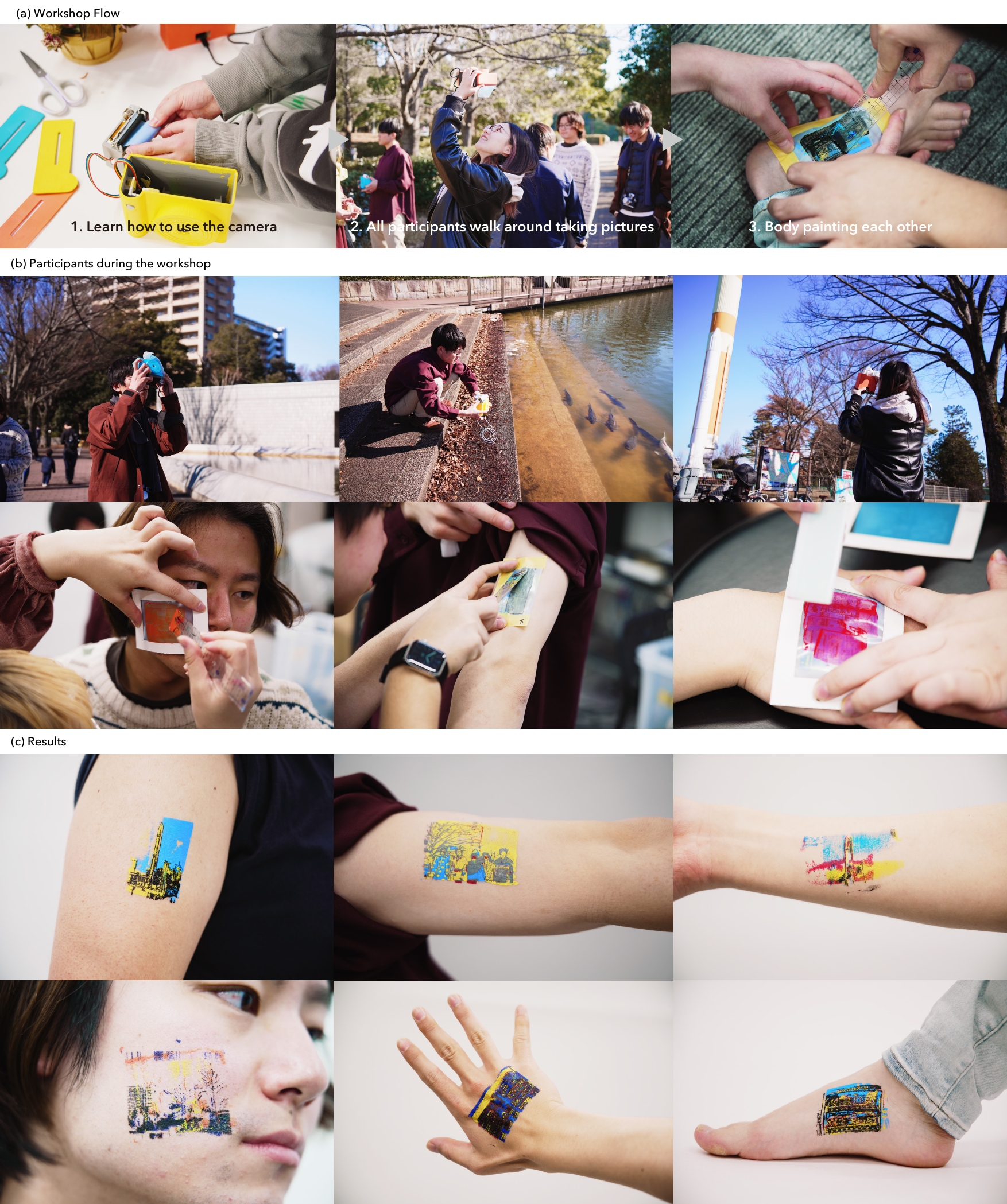}
    \caption{(a) Workshop overview. (b) Participants took photos outdoors in a park and science museum, then performed multicolor printing indoors. (c) Six participants body-painting each other, applying prints to areas like arms, hands, foot, and cheek.}
    \label{fig:ws-results}
\end{figure}

The purpose of the workshop was to experientially understand the relationship between photographic technology and body ornamentation by applying photographic body paint on their bodies. The flow is shown in Figure\ref{fig:ws-results}(a).
First, instructions were provided on how to use the camera and paint on the body. Next, participants engaged in free shooting both indoors and outdoors using our instant camera. The photographs taken were printed onto mesh screens and applied to each participant's body using silkscreen techniques. Three cameras were prepared and shared among the participants. The total duration was three hours, including breaks. After the body painting, participants photographed the painted areas and answered a questionnaire. Four of them joined semi-structured interviews at one's will.

\subsection{Results and Findings}

\subsubsection{Results}

Participants experienced the process wherein digital photographs obtained physical substance through a mesh screen and melted into their bodies as body ornamentation. They became accustomed to using the camera, exploring appropriate shooting methods while checking the printed images, and collaborating in painting each other's bodies. Natural interactions emerged even among those meeting for the first time.
By directly printing photographs onto the body, new appreciative experiences and communication emerged. Photography acquired corporeality, expanding its role as a medium. For example, the situation of painting one's own body with a photograph taken by someone else became an attempt to reconstruct the self from another's perspective and deepen relationships. Additionally, by photographing others and painting those images onto one's own body, expressions that blurred the boundaries between self and other, and between inside and outside, were observed.
The fusion of photography and body ornamentation gave rise to new forms of expression. From the perspective of McLuhan's assertion that "the medium is the message," it can be interpreted that the medium itself is generating new meanings.

\subsubsection{Unexpected and Impressive Behaviors}

\begin{figure}[htbp]
    \centering
    \includegraphics[width=0.9\linewidth]{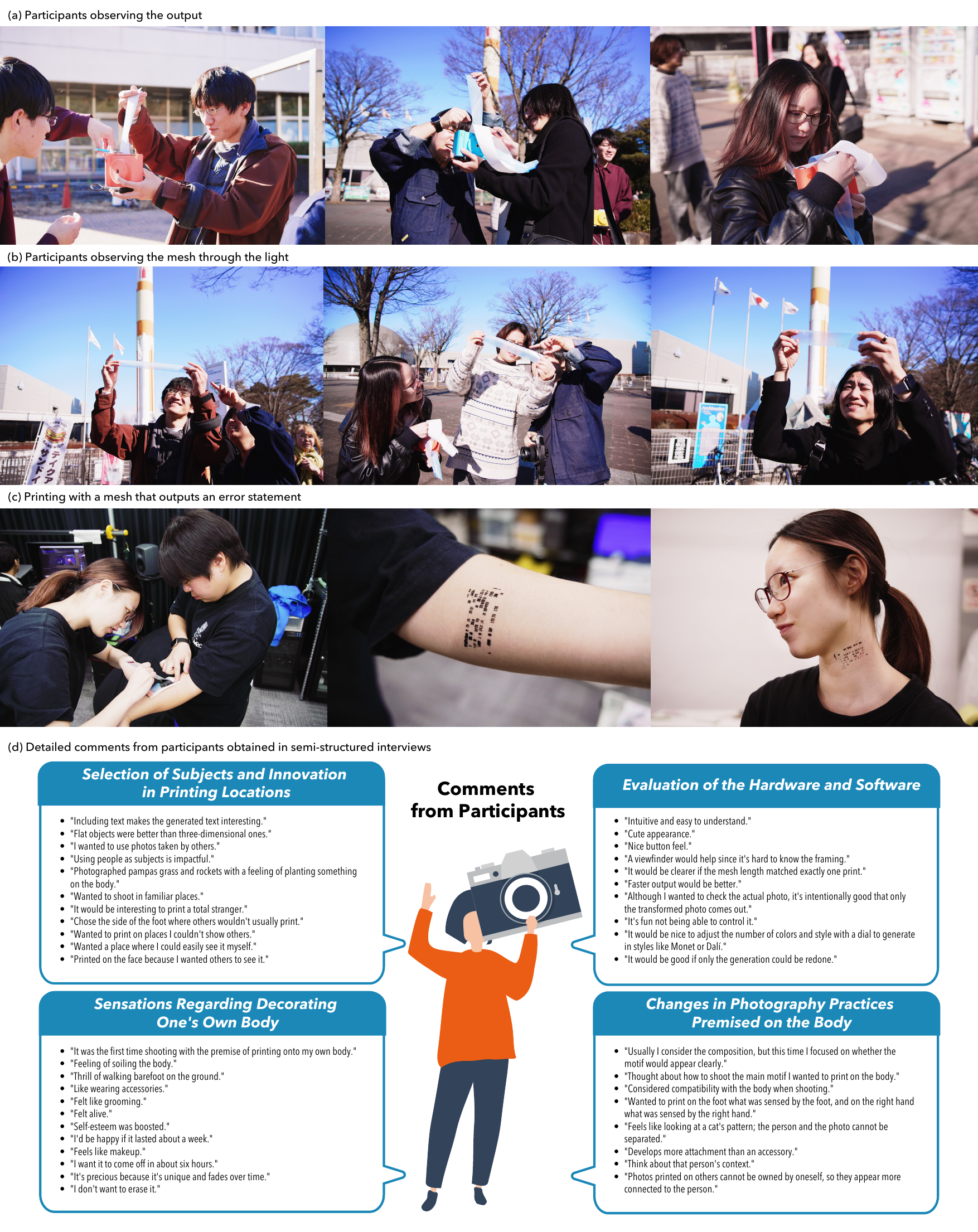}
    \caption{(a) Participants observe mesh output from the camera, collaborating on tasks like cutting and replenishing. (b) Examining the mesh held up to light, like viewing negatives. (c) A participant body-painting using a mesh with error codes. (d) Participants' comments from interviews on subject selection, printing locations, feelings about body ornamentation, hardware and software evaluations, and changes in photographic practices.}
    \label{fig:ws-findings}
\end{figure}

Participants eagerly awaited the moment when the mesh screen was output from the camera, and they shared that moment with cheers upon its release(Figure\ref{fig:ws-findings}(a)). This process was one of the elements that invigorated communication. The sense of unity of sharing the waiting time is a pleasure that digital cameras lack. The output process enhanced participants' expectations, conversations, and emotional sharing, and created value as a collaborative experience.

Moreover, participants showed interest in the texture and transparency of the mesh screens(Figure\ref{fig:ws-findings}(b)). Holding them to the light to observe the image like observing film negatives, brings out a fresh surprise toward analog experiences. Furthermore, when error codes were output due to camera malfunctions, participants did not regard them as failures but embraced them as unique designs for body painting(Figure\ref{fig:ws-findings}(c)). This attitude of having unexpected results as creative opportunities reflects a reevaluation of contingency value.

Consequently, the participants' unexpected behaviors and reactions led to a renewed recognition of the new expressive possibilities of the fusion of photographic technology and body ornamentation. It also underscored the importance of the materiality and processes of media.

\subsection{Results and Discoveries from the Questionnaire and Semi-Structured Interviews}

After the workshop, semi-structured interviews were conducted with four participants to explore their experiences and impressions in detail. Detailed comments are presented in Figure\ref{fig:ws-findings}(d). All participants expressed satisfaction, stating that they enjoyed the experience. Specific comments included that reprinting and printing on their skin lifted their mood, the time allocation felt perfect, and the painting was fun but also challenging. They particularly enjoyed the moment of checking the print after the shooting. The instant the output appeared was a mix of expectation and surprise; some remarked that they felt disappointed when a void was printed. Additionally, they enjoyed the creative process and interaction with others, noting that thinking about what to print where, painting on others, being painted by others, and creative aspects such as choosing colors were enjoyable.

Participants demonstrated unique perspectives and ingenuity in selecting subjects and body parts for printing. The actions of painting the body provided a fresh sensation, prompting changes in self-expression and awareness of corporeality. Some participants noted that positive effects were gained through interactions with others, deepening communication.

Regarding our cameras, while participants found them intuitive and easy to use, they also suggested improvements and additional features. Through photographic activities premised on printing onto the body, participants think of perspectives and creativity different from regular photography.

These results discovered that the novel expression by fusing photographic technology and body ornamentation provided participants with profound experiences and diverse discoveries. It offers meaningful insights for exploring the materiality of media and the possibilities of expression through the body.

\section{Discussion and Conclusion}

\subsection{Discussion on the Boundary Between Tattoos and Body Paint}

\begin{figure}[htbp]
    \centering
    \includegraphics[width=1.0\linewidth]{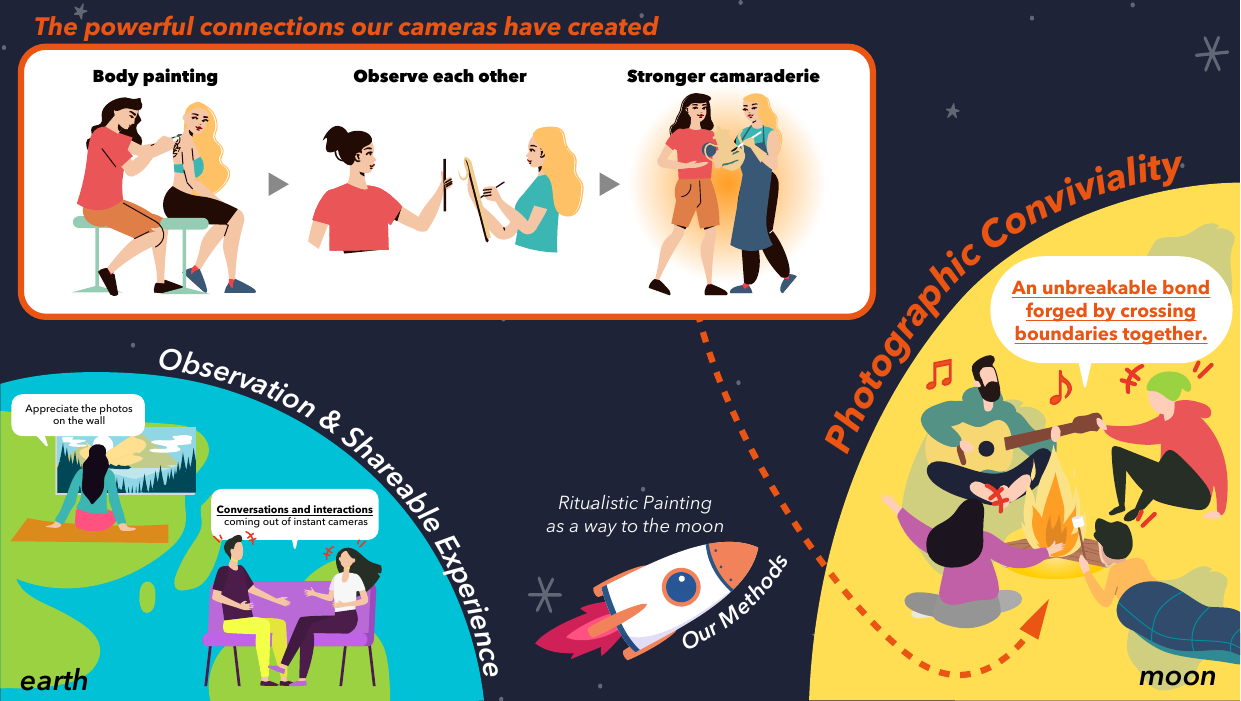}
    \caption{Design space of Photographic Conviviality: Shifting from familiar static photographs ("Earth") to dynamic photographs ("Moon") represents a courageous venture into new expressive territories.}
    \label{fig:discussion}
\end{figure}

Since ancient times, tattoos have functioned as symbols indicating ethnic and social identity, fulfilling ritual roles such as initiation into social groups or classes. In contrast, body painting, with its variability and immediacy, enables personal self-expression and temporary artistic creativity. Exploring the intersection of these two forms of expression has revealed new possibilities and social significance in body ornamentation.
Interviews with workshop participants revealed statements such as "It feels like self-harm" and "I felt like I was defiling someone else's body," indicating the presence of psychological resistance to applying body paint to oneself or others. This resistance likely reflects deep-seated social and cultural norms, as well as personal internal conflicts regarding direct intervention and transformation of the body. However, within the workshop setting, when one participant bravely painted the body, others were inspired by the act and its aesthetic outcome, leading to positive shifts in attitude, such as admiring, "I want to try it too!"
This phenomenon suggests that the temporality and variability of body painting play a role in fading psychological resistance through conviviality and shared experiences. The characteristic that body painting is temporary is a ritual that is valid only for its duration. Participants who pass through this process naturally become friends with each other, strengthening a sense of unity and empathy. Furthermore, once this psychological hurdle is overcome, participants' resistance to body painting rapidly fades, releasing mental and physical strain tension. Such contemporaneous empathy and heightened energy are unique attractions of body expression and significant social value.
Reflecting on the history of photographic technology, instant cameras like Polaroid and Instax have revitalized communication and the sharing of memories by producing photographs here. However, the instant camera developed in this study offers a new dimension of experience by not only allowing visual appreciation of photographs but also melting images into the body. Through this process, photographs expand from being visual recording media to becoming a body of self-expression and interaction with others.
In the workshop, participants printing photographs taken by each other onto their bodies created a more dynamic and celebratory space that transcended conventional photo appreciation and sharing. We defined this phenomenon as "Photographic Conviviality", a concept referring to the intimacy, solidarity, and deepening of shared experiences that arise among people mediated by photography.

\subsection{Conclusion}

In this paper, possibilities for expression and communication are explored through the fusion of photographic technology and body ornamentation, identifying common ground between permanent tattoos and temporary body painting.

Through the workshop, participants experienced new forms of self-expression by incorporating photographs into their bodies. By directly printing photographs onto their skin, these images became part of their physical and linguistic expression, functioning as an extension of self-expression.

This process transformed photographs from static, two-dimensional images into dynamic bodily experiences. Participants used their bodies as canvases, reinterpreting pictures' documentary nature and reconstructing them as more subjective means of self-expression. This reevaluation of photography's aura presents new relationships among photography, the body, and self-expression in the digital age.

Future plans include workshops with diverse participants to examine differences in reactions and experiences due to cultural backgrounds and age groups. Additionally, technical improvements are expected to enable high-definition printing and diverse methods of expression, further expanding the artistic possibilities brought about by the fusion of photography and body ornamentation.

Through this research, the concept of "photographic conviviality" was proposed, demonstrating that the fusion of photography and body ornamentation creates new celebratory forms of communication. It provides important insights into rethinking self-expression and relationships with others in contemporary society, indicating that the joy of co-creating new narratives through the body can be shared.

\bibliographystyle{ACM-Reference-Format}
\bibliography{sample-base}

\end{document}